\documentclass[sigconf]{acmart}
\usepackage{bm}
\usepackage[ruled, lined, longend, linesnumbered ]{algorithm2e}

\AtBeginDocument{%
  \providecommand\BibTeX{{%
    \normalfont B\kern-0.5em{\scshape i\kern-0.25em b}\kern-0.8em\TeX}}}

\copyrightyear{2021} 
\acmYear{2021} 
\setcopyright{acmlicensed}\acmConference[ICAIF'21]{2nd ACM International Conference on AI in Finance}{November 3--5, 2021}{Virtual Event, USA}
\acmBooktitle{2nd ACM International Conference on AI in Finance (ICAIF'21), November 3--5, 2021, Virtual Event, USA}
\acmPrice{15.00}
\acmDOI{10.1145/3490354.3494376}
\acmISBN{978-1-4503-9148-1/21/11}

\begin{document}

\title{An Automated Portfolio Trading System with Feature Preprocessing and Recurrent Reinforcement Learning}
\author{Lin Li}
\affiliation{\institution{Southern University of Science and Technology} \city{Shenzhen} \country{China} }
\email{linleeccnu@gmail.com}



\begin{abstract}
We propose a novel portfolio trading system, which contains a feature preprocessing module and a trading module. The feature preprocessing module consists of various data processing operations, while in the trading part, we integrate the portfolio weight rebalance function with the trading algorithm and make the trading system fully automated and suitable for individual investors, holding a handful of stocks. The data preprocessing procedures are applied to remove the white noise in the raw data set and uncover the general pattern underlying the data set before the processed feature set is inputted into the trading algorithm. Our empirical results reveal that the proposed portfolio trading system can efficiently earn high profit and maintain a relatively low drawdown, which clearly outperforms other portfolio trading strategies.   
\end{abstract}



\maketitle

\section{Introduction}
Portfolio trading usually aims to maximize the return over time and minimize the investment risk simultaneously. Investors typically gain profits by dynamically allocating their wealth among selected assets at the initial period and rebalancing their wealth afterwards. 
With the fast development of machine learning in recent years, portfolio trading has been extensively studied in the machine learning community. \cite{moody1997optimization,moody1998performance} proposed to use recurrent reinforcement learning (RRL) algorithms to optimize trading systems and the algorithm was extended to optimize portfolios consisting of the S\&P 500 Stock Index and T-Bills \cite{moody2001learning}. Afterwards, other researchers followed their scheme with certain variations. \cite{hens2007strategic} applied RRL to the strategic asset allocation on samples from various countries. They only consider a portfolio consisting of 2 assets, long-term bond and equity and let the portfolio weight of one asset $w_t$ depend on the parameters of the RRL algorithm by a heuristic equation in the economic context.  \cite{maringer2012regime} 
constructed a portfolio consisting of 12 stocks and let the position of each stock, long (1) or short (-1), be decided by the trading signal from each stock trading system, which was driven by the proposed regime-switching RRL. Also, \cite{almahdi2017adaptive} combined the RRL trading algorithm with the Calmar ratio to do the portfolio rebalance. They used the RRL algorithm to optimize the Calmar ratio instead of Sharpe ratio to generate the trading signals or positions of each asset, then the portfolio weight was decided by an exogenous softmax function based on the trading signal of each asset. On the other hand, online learning has also been broadly applied to select portfolios. \cite{li2012pamr} proposed passive aggressive mean reversion strategy, which constructed portfolios to minimize the deviation from the last portfolio. \cite{li2013confidence} exploited the mean reversion and variance information of portfolios and proposed the confidence weighted mean reversion strategy. Additionally, \cite{li2015moving} proposed the on-line portfolio selection with moving average reversion (OLMAR) strategy, which used the moving average reversion pattern of stock price relatives. 
A notable difference between the portfolio selection algorithm driven by RRL and online learning is that online learning techniques allow more assets to be included in a portfolio than the reinforcement learning algorithm does. Other machine learning techniques and models have also been applied in portfolio selection. For instance, \cite{2021Mean} developed a novel portfolio selection approach using a machine learning algorithm, eXtreme Gradient Boosting with an improved firely algorithm, and the mean-variance portfolio choice theory \cite{markowitz1952portfolio}. A mixed portfolio formation approach was proposed in \cite{2020Portfolio}, which used the long-short term memory networks to preselect stocks with high potential return and then the mean-variance model was applied to do the portfolio construction with the chosen stocks. Additionally, support vector machine and mean-variance model were combined to select portfolios in \cite{2018Decision}. 

Amongst these machine learning portfolio selection models, we are particularly interested in RRL models, because it lays the ground for automated trading where
the trading signals of each asset over the investment horizon are directly generated and its characteristics make it more suitable for individual investors. However, all these previous work about portfolio selection or trading mainly focus on building various machine learning algorithms to boost the model performance. To the best of our knowledge, few work has been dedicated to improving the feature quality before implementing the trading algorithms. In this paper, we build an automated portfolio selection system, named PCA\&DWT RRL, based on RRL and two main data preprocessing approaches, Principal Component Analysis (PCA) and Discrete Wavelet Transform (DWT). The PCA\&DWT RRL system is driven by the RRL algorithm that aims to maximize the risk-adjusted return and contains some auxiliary components for feature processing. Specifically, the data set containing the daily close price and volume from each stock is firstly preprocessed to extract various technical indicators which are then further preprocessed by PCA and DWT techniques to form the feature series for each stock. Then the daily feature streams of various indicators are combined into one dataframe for each stock. Then for each period, the daily feature vectors of various stocks are concatenated as one element of the final feature set which is inputted into the RRL algorithm to generate portfolio weights at each trading period.

Our contributions are: 1, we embed the portfolio rebalance function into the RRL trading algorithm and the adjustment of portfolio weight is internally automated, which differentiates our approach from other portfolio trading methods based on RRL. 2, we integrate several data preprocessing steps with the trading algorithm, which ensures a high quality of the feature set for trading. 3, we have done extensive experiments to test the performance of PCA\&DWT RRL system and the empirical results demonstrate that the proposed system consistently outperforms other benchmark portfolio selection strategies. Moreover, we find that feature preprocessing plays a vital role in this setting.  


\section{Data Preprocessing Module}

\subsection{Data Configuration Layer}

In the financial market, the price of stocks is affected by multiple factors which include but not limited to unexpected events, government policies and company activities. This implies that there may be a few periods when trading activities are not available, which results in a couple of NA entries in the price time series. These NAs are useless for our trading system and can be removed directly or filled out with other values subject to specific learning task. In our case, we discard these NAs directly for simplicity. Note that removing NA entries may result in irregular samples in timeline for different stocks. Hence, here we make sure the length of sample points is the same for all stocks and assume they are of the same timeline.   


Technical analysis in the financial market has a long history amongst investment practitioners. However, it was omitted by academics over a few decades in the past probably due to the belief of efficient market hypothesis. As more and more evidence shows that markets are not as efficient as once believed. Technical analysis were applied to algorithmic trading \cite{dempster2001computational,nobre2019combining}. It is generally believed that technical analysis indicators can summarize the general pattern of the time series and avoid local noise in the data stream, which can further be utilized by the trading system to make profitable decisions. There are  quite  a  few  technical indicators developed by financial professionals \cite{achelis2001technical}. One can choose various technical indicators to use depending on specific tasks. While without enough number of indicators, it may be tough to reveal the pattern of the data stream comprehensively, including too many technical indicators may also affect the trading decision negatively and increase the computation burden, since the calculated value of some technical indicators are not always consistent with one another. Given this, the indicators selected in this paper can be categorized into four groups with the hope to reflect the overall history information of specific stocks, momentum indicators: Momentum (MOM), Moving Average Convergence Divergence (MACD), Money Flow Index (MFI), Relative Strength Index (RSI), volatility indicators: Average True Range (ATR), Normalized Average True Range (NATR), cycle indicators: Hilbert Transform Dominant Cycle Phase (HTDCP), Hilbert Transform Sinewave (HTS), Hilbert Transform Trend Market Mode (HTTMM) and volume indicators: Chaikin Oscillator (CO), On Balance Volume (OBV). We select these technical indicators due to Let $TA$ = \{MOM, MACD, MFI, RSI, ATR, NATR, HTDCP, HTS, HTTMM, CO, OBV\} denote the set of 11 technical indicators used for the following exposition.  
The calculation of technical indicators was done via the python library TA-Lib \cite{talib}.

Additionally, to make the value of each feature on the same scale, we normalize each
technical indicator stream with the standard normalization, i.e. z-score to process each technical indicator series, which is represented by the following equation.
\begin{equation}
    X = \frac{X-\mu(X)}{\sigma(X)}
\end{equation}
where $X$ is the time series of each extracted feature, $\mu(X)$ and $\sigma(X)$ is the mean and standard deviation of $X$, respectively.

\subsection{Principal Component Analysis Layer} 
The PCA layer of data preprocessing module receives the normalized feature set, including 11 technical indicators data streams, as input. The development of PCA originates from the curse of dimensionality which claims that data points in high dimension lie far away from each other, statistically speaking \cite{geron2019hands}. The curse of dimensionality not only makes the training of machine learning algorithm expensive, but also casts a shadow over the predictions of the trained algorithm, since data points in-sample and out-of-sample are so far away. In order to alleviate the effect of curse of dimensionality, a natural way is to reduce the dimension of data sets. That is where PCA comes into effect. To be more specific, PCA firstly fits the input, identifying the main components that represent the directions of maximum variance of the input. Then the main components are ordered according to the variance they explain and one can choose how many components are preserved. Afterwards, the original input is projected onto the retained components, resulting in a data set of lower dimension. In our case, the normalized technical indicators is a 11 dimensional data set at first. After being decomposed by PCA, there will be less than 11 indicators, reducing the probability of correlation and inconsistency amongst different technical indicators. In this way, the processed $TA$ set results in a new $TA^\prime \subset TA$. Note that different stocks may have different components of technical indicators retained in $TA^\prime$. Here we only need to make sure that the cardinality of $TA^\prime$s is the same for all stocks to meet the logic of the RRL trading algorithm afterwards. 
We use the well developed Scikit-learn \cite{pedregosa2011scikit} library to implement the PCA operation. Additionally, we set the hyperparameter, explained variance ratio, to 95\% which means that the sum of the variance explained by all retained principal components takes up at least 95\% of the total variance of the original data set.

\subsection{Discrete Wavelet Transform Layer}
Although the data set returned by the PCA layer is simplified by removing less relevant features in the feature domain, some outliers or irrelevant data points, representing local noise, may still exist in each feature series. They may affect the training and trading of the RRL algorithm. To remove these local noise in the time domain of each indicator, we apply the discrete wavelet transform to the feature data after being processed by PCA.

Reference \cite{mallat1989theory} proposed to calculate the DWT coefficients, including the approximation and the detail coefficients, using a pair of high pass and low-pass filter. In the algorithm, the father $\Phi(t)$ and mother $\Psi(t)$ basis functions are introduced to generate their corresponding son $\Phi_{j,k}(t)$ and daughter $\Psi_{j,k}(t)$ wavelets which are further utilized to approximate the original signal. The corresponding coefficients of son and daughter wavelets as a result of decomposing a function $f(x)$ is defined in the following inner product form \cite{mallat1989theory}. 
\begin{equation}
a_{j,k} = \langle f(t), \Phi_{j,k}(t) \rangle, \quad
d_{j,k} = \langle f(t), \Psi_{j,k}(t) \rangle
\end{equation}
where $a_{j,k}$ and $d_{j,k}$ are approximation and detail coefficients, respectively, and $k=0,1,2,\ldots$ and $j = 0,1,2,\ldots$. Though there are various types of wavelets, in this work, we use Haar wavelets with periodization padding mode, since Haar wavelets are useful to capture fluctuations between adjacent observations, recorded by \cite{Lahmiri2014Wavelet}, which would be heuristically useful to spot evident drawdowns in the financial market. Additionally, the decomposition can be iterated for multiple times, subject to the inherent decomposition level of the wavelets and the length of the signal or data series. 

Eventually, the DWT leaves us one set of the approximation coefficients and a couple of sets of the detail coefficients depending on the given decomposition level. In this paper, we set the DWT decomposition level equal to 4, since too high decomposition level would destroy the general pattern, while too low would still leave too much noise in the data \cite{lee2019pywavelets}. After the decomposition finishes, the general trend of the original signal is preserved in the approximation set, while the detail coefficients sets contain the local noise of the signal \cite{mallat1989theory}, which we aim to clean. At this point, we apply soft thresholding technique with an empirical threshold value equal to two times standard deviations of coefficients to each detail coefficients set. Eventually, the inverse DWT method is used to reconstruct the signal which is the final denoised version of the original signal. By discarding the irrelevant coefficients, the reconstructed signal represents the essential characteristics of the original signal, which is further adopted by the RRL trader in the following trading module. The DWT process for each technical indicator series of each asset is implemented with the open source python package PyWavelets \cite{lee2019pywavelets}. Note that here parameters for DWT are kept the same for the PCA-transformed technical indicators.

\section{Recurrent Reinforcement Learning for Portfolio Trading Module}
\subsection{Portfolio Rebalance Function}
In this paper, we assume that the trader takes only long positions and there is no income or consumptions. 
At the beginning of each period, the trader rebalances the portfolio which is composed of several securities with corresponding weights. Assuming there are $m$ securities with price series $\{\{p_t^a\}: a=1,...,m\}$, the market rate of return $r_t^a$ for price series $p_t^a$ for the period ending at time $t$ is defined as $r_t^a = \frac{p_t^a}{p_{t-1}^a} - 1 $ and thus the return vector of $m$ securities is defined as $\bm{r}_t=[r_t^1, r_t^2, ..., r_t^m]^\top$. Defining portfolio weight of the $a^{th}$ security at period $t$ as $F_t^a$, $\bm{F}_t = [F_t^1, F_t^2, ..., F_t^m]^\top$ and the vector $\bm{1}=[1,1,...,1]^\top$, then the trader that takes only long positions must have portfolio weights that satisfy:
\begin{equation}
\bm{F}_t^\top \cdot \bm{1}  =1, \quad \bm{F}_t  \geq 0
\end{equation}
Given these conditions, we use the following normalized outputs as portfolio weights:
\begin{equation}\label{e3}
\bm{F}_t = \frac{\text{exp}[\bm{f}_t(\bm{Y}_t)]}{\bm{1}^\top\cdot \text{exp}[\bm{f}_t(\bm{Y}_t)]} 
\end{equation}
which is suggested by \cite{moody1997optimization,moody1998performance} and $\bm{f}_t$ is defined using hyperbolic tangent activation function as:
\begin{align}\label{key}
\bm{f}_t(\bm{Y}_t) & = \text{tanh}(\bm{Y}_t) \\
\bm{Y}_t & = (\bm{X}_t \otimes \bm{\Theta})\cdot \bm{1}
\end{align}
where $\bm{X}_t = [\bm{x}_t^1,\bm{x}_t^2,...,\bm{x}_t^m]^\top $ is the input feature matrix to the trading system, while $\bm{\Theta} = [\bm{\theta}^1,\bm{\theta}^2,...,\bm{\theta}^m]^\top $ is the system parameter matrix to be learned during the training process, and $\otimes$ represents the element-wise product between matrices. Note that all elements of matrices $\bm{X}_t$ and $\bm{\Theta}$ are vectors per se, in particular, $\bm{x}_t^a=[1,ta_t^1,ta_t^2,\ldots,ta_t^n,F_{t-1}^a]$, where $ta_t^1,ta_t^2\ldots,ta_t^n$ are the value of technical indicators remained in $TA^\prime$ at period $t$ for security $a$ and $n$ is the cardinality of the $TA^\prime$ set.


\subsection{Profit of Portfolio Tradings}

Since $F_t^a$ represents the holdings of security $a$ at period $t$, then $F_t^a$ should be re-adjusted at each time step according to Equation \ref{e3}. Thus, generally speaking, a transaction cost rate $\delta$ should be applied to each security weight adjustment between two consecutive periods. 
One can arguably expect that higher transaction costs would discourage the excessive rebalance actions \cite{moody2001learning}. Since our focus is the systematic construction of the trading system, we here assume the influence of price series movements on portfolio weights is negligible for the ease of exposition and analysis. In this case, the wealth of the trading at time $T$ is:
\begin{equation}\label{eq8}
\begin{split}
W_T & = W_0 \prod_{t=1}^T (1 + R_t) \\
& = W_0 \prod_{t=1}^T (1 + \bm{F}_{t-1}^\top\bm{r}_t )(1 - \delta\cdot \bm{1}^\top|\bm{F}_t - \bm{F}_{t-1}|)
\end{split}
\end{equation} 
where $W_0$ is the initial wealth of the investment account, which we set as \$1 for simplicity, and $R_t$ is the return of the portfolio at time $t$ i.e. one-stage profit, defined as: 
\begin{equation}\label{eq9}
R_t =  (1 + \bm{F}_{t-1}^\top\bm{r}_t )(1 - \delta\cdot \bm{1}^\top|\bm{F}_t - \bm{F}_{t-1}|) - 1
\end{equation}
In this case, the cumulative profit obtained from the investment after $T$ periods is:
\begin{equation}\label{key}
P_T = W_T - W_0
\end{equation}

\subsection{Sharpe Ratio}
In this paper, we aim to optimize the risk-adjusted return of the portfolio (e.g. Sharpe ratio) among other performance criteria, since it is relatively simple and widely adopted by various investors.
The Sharpe ratio is commonly defined as the ratio between the average and standard deviation of a period of historical returns, $R_{1,\ldots,T}$ \cite{moody2001learning},
where $R_t$ is the return of investment at trading period $t$.
Intuitively, Sharpe ratio rewards investment strategies that are less volatile to make profits.

\subsection{Gradient Ascent}
To obtain the optimal portfolio rebalance strategy, the RRL algorithm needs to learn the optimal parameters via maximizing Sharpe ratio of the portfolio. Therefore, one needs to evaluate the influence of Sharpe ratio on the portfolio trading system during training. We attain this goal by computing the first order derivative of Sharpe ratio with respect to (w.r.t.) $\bm{\Theta}$. Furthermore, we adopt the gradient ascent to update the model parameters learned during training. Although automatic differentiation is easily available, we include technical details here for the reference of implementation, especially for readers with little background in machine learning. 

First of all, with the estimate of the first and second moments of returns distributions, one has the Sharpe ratio formula of a portfolio as follows \cite{moody2001learning}:
\begin{equation}\label{key}
\begin{split}
S_T & = \frac{E[R_{1,\ldots,T}]}{\sqrt{E[R_{1,\ldots,T}^2]-(E[R_{1,\ldots,T}])^2}} = \frac{A}{\sqrt{B-A^2}} \\
\end{split}
\end{equation}
Where $A = \frac{1}{T}\sum_{t=1}^{T}R_t$, $B =\frac{1}{T}\sum_{t=1}^{T}(R_t)^2$ and $R_t$ is the return of the portfolio at time $t$. Then, the first order derivative of $S_T$ w.r.t. the system parameters is computed using the chain rule:
\begin{equation}\label{13}
\begin{split}
\frac{dS_T}{d\bm{\Theta}} & = \frac{d}{d\bm{\Theta}}\bigg\{\frac{A}{\sqrt{B-A^2}}\bigg\}   \\
& = \frac{\partial S_T}{\partial A}\cdot\frac{\partial A}{\partial \bm{\Theta}}+\frac{\partial S_T}{\partial B}\cdot\frac{\partial B}{\partial \bm{\Theta}} \\
& = \sum_{t=1}^{T}\bigg\{ \frac{\partial S_T}{\partial A}\cdot\frac{\partial A}{\partial R_t}+\frac{\partial S_T}{\partial B}\cdot\frac{\partial B}{\partial R_t} \bigg\} \cdot \frac{dR_t}{d\bm{\Theta}} \\
& = \sum_{t=1}^{T}\bigg\{ \frac{\partial S_T}{\partial A}\cdot\frac{\partial A}{\partial R_t}+\frac{\partial S_T}{\partial B}\cdot\frac{\partial B}{\partial R_t} \bigg\} \\
& \cdot \bigg\{ \text{diag}(\frac{\partial R_t}{\partial \bm{F}_t})\frac{\partial \bm{F}_t}{\partial \bm{\Theta}}+ \text{diag}(\frac{\partial R_t}{\partial \bm{F}_{t-1}})\frac{\partial \bm{F}_{t-1}}{\partial\bm{\Theta}} \bigg\}
\end{split}
\end{equation}
where \text{diag}$(\frac{\partial R_t}{\partial \bm{F}_t})$ and \text{diag}$(\frac{\partial R_t}{\partial \bm{F}_{t-1}})$ stand for square matrices whose main diagonal entries are from vectors $\frac{\partial R_t}{\partial \bm{F}_t}$ and $\frac{\partial R_t}{\partial \bm{F}_{t-1}}$, respectively, and all other entries are 0. Note that the difference between the direct portfolio optimization as our method and single security automated trading optimization \cite{moody2001learning} 
is that here $\frac{dS_T}{d\bm{\Theta}}$ is no longer a vector, but a matrix. Since we are trading several risky assets simultaneously, this means $\bm{F}_t$ and $ \bm{F}_{t-1}$ are both vectors, and $\bm{\Theta}$ is a matrix. The Jacobian matrices should be calculated for partial derivatives, $\frac{\partial R_t}{\partial \bm{F}_t}, \frac{\partial \bm{F}_t}{\partial \bm{\Theta}}, \frac{\partial R_t}{\partial \bm{F}_{t-1}}$ and $\frac{\partial \bm{F}_{t-1}}{\partial\bm{\Theta}}$. To be more specific:
\begin{equation}\label{key}
\begin{split}
\frac{\partial R_t}{\partial \bm{F}_t} = &  \frac{\partial}{\partial \bm{F}_t}\big\{(1 + \bm{F}_{t-1}^\top \bm{r}_t )(1 - \delta\cdot \bm{1}^\top|\bm{F}_t - \bm{F}_{t-1}|) - 1 \big\} \\
= & - \delta \cdot (1 + \bm{F}_{t-1}^\top \bm{r}_t)\cdot \text{sgn}(\bm{F}_t-\bm{F}_{t-1})
\end{split}
\end{equation}
\begin{equation}\label{key}
\begin{split}
\frac{\partial R_t}{\partial \bm{F}_{t-1}}  = &  (1 - \delta\cdot \bm{1}^\top|\bm{F}_t - \bm{F}_{t-1}|)\bm{r}_t  \\
& + \delta \cdot (1 + \bm{F}_{t-1}^\top \bm{r}_t)\cdot \text{sgn}(\bm{F}_t-\bm{F}_{t-1})
\end{split}
\end{equation}
\begin{equation}\label{key}
\begin{split}
\frac{\partial \bm{F}_t}{\partial \bm{\Theta}}  = &{} \frac{\partial \bm{F}_t}{\partial \bm{f}_t}\cdot \frac{\partial \bm{f}_t}{\partial \bm{Y}_t}\cdot (\frac{\partial \bm{Y}_t}{\partial \bm{\Theta}} + \frac{\partial \bm{Y}_t}{\partial \bm{F}_{t-1}} \cdot \frac{\partial \bm{F}_{t-1}}{\partial \bm{\Theta}})\\
= & \bm{D}\bm{F}_t(\bm{f}_t)\cdot \bm{D}\bm{f}_t(\bm{Y}_t)\cdot (\bm{X}_t + \text{diag}(\bm{\theta}^{\prime\top}) \frac{\partial \bm{F}_{t-1}}{\partial \bm{\Theta}}) 
\end{split}
\end{equation}
Note that here we choose $\bm{f}_t$ in the form of \text{tanh} function, the logistic function form of $\bm{f}_t$ can also be easily calculated. Moreover, $\bm{D}\bm{F}_t(\bm{f}_t)$ and $\bm{D}\bm{f}_t(\bm{Y}_t)$ are the Jacobian matrices of $\bm{F}_t$ w.r.t. $\bm{f}_t$ and $\bm{f}_t$ w.r.t. $\bm{Y}_t$, respectively, and $\bm{\theta}^\prime$ is the vector of the last column of parameter matrix $\bm{\Theta}$, corresponding to $\bm{F}_{t-1}$ in the feature matrix, $\bm{X}_t$.
\begin{equation}\label{key}
\bm{D}\bm{F}_t(\bm{f}_t) =
\begin{bmatrix}
\frac{\partial F_t^1}{\partial f_t^1} & \frac{\partial F_t^1}{\partial f_t^2} & \dots & \frac{\partial F_t^1}{\partial f_t^m} \\
\frac{\partial F_t^2}{\partial f_t^2} & \ddots & \frac{\partial F_t^i}{\partial f_t^j} & \dots \\
\vdots & \vdots & \vdots & \vdots \\
\frac{\partial F_t^m}{\partial f_t^1} & \frac{\partial F_t^m}{\partial f_t^2} & \dots & \frac{\partial F_t^m}{\partial f_t^m}
\end{bmatrix}
\end{equation}
Define $S_i = \frac{\text{exp}(f_t^i)}{\sum_i\text{exp}(f_t^i)}$, then entries of the Jacobian $\bm{D}\bm{F}_t(\bm{f}_t)$ can be simplified as:
\begin{equation}\label{key}
\frac{\partial F_t^i}{\partial f_t^j} = 
\begin{cases}
S_i(1-S_j)  & \text{if} \quad i=j  \\
-S_iS_j &  \text{if} \quad i \neq j
\end{cases}
\end{equation}
Similarly, the Jacobian $\bm{D}\bm{f}_t(\bm{Y}_t)$ is calculated as follows:
\begin{equation}\label{key}
\bm{D}\bm{f}_t(\bm{Y}_t) =
\begin{bmatrix}
\frac{\partial f_t^1}{\partial Y_t^1} & 0 & \dots & 0 \\
0 &  \frac{\partial f_t^2}{\partial Y_t^2} & \ddots  & 0 \\
\vdots & \vdots & \vdots & \vdots \\
0 & 0 & \dots & \frac{\partial f_t^m}{\partial Y_t^m}
\end{bmatrix}
\end{equation}
where
\begin{equation}\label{key}
\frac{\partial f_t^i}{\partial Y_t^i} = 1 - \text{tanh}^2(Y_t^i) 
\end{equation}
It is straightforward that the derivative $\frac{\partial \bm{F}_t}{\partial \bm{\Theta}}$ is recurrent and depends on all its previous value within the time window $T$. This is also the reason why this reinforcement learning method is recurrent. Once the term $\frac{dS_T}{d\bm{\Theta}}$ has been calculated, the system parameters $\bm{\Theta}$ is updated according to the gradient ascent rule with
consideration of the $\ell_2$ regularization to avoid overfitting the noise in the data, 
\begin{equation}\label{key}
\begin{split}
\bm{\Theta}_{n+1} & = \bm{\Theta}_n + \rho \cdot (\frac{dS_T}{d\bm{\Theta}_n} - \lambda \cdot \bm{\Theta}_n) \\
& = (1-\rho\cdot\lambda)\cdot\bm{\Theta}_n + \rho\cdot\frac{dS_T}{d\bm{\Theta}_n}
\end{split}
\end{equation}
where $\rho, \lambda \geq 0$ are the given learning rate and $\ell_2$ regularization hyperparameter. The process is repeated for $N$ epochs, where $N$ can be chosen such that Sharpe ratio has converged during training.
Note that optimizing Sharpe ratio requires to calculate the gradient of Sharpe ratio w.r.t. $\bm{\Theta}$ which depends on the total derivative of $\frac{\partial \bm{F}_t}{\partial \bm{\Theta}}$. Therefore, we adopt an efficient recurrent algorithm similar to backpropagation through time (BPTT) to train the model, as in \cite{moody2001learning} for single stock trading using RRL.

\subsection{Algorithm}
\begin{figure}
  \centering
  \includegraphics[width=\columnwidth]{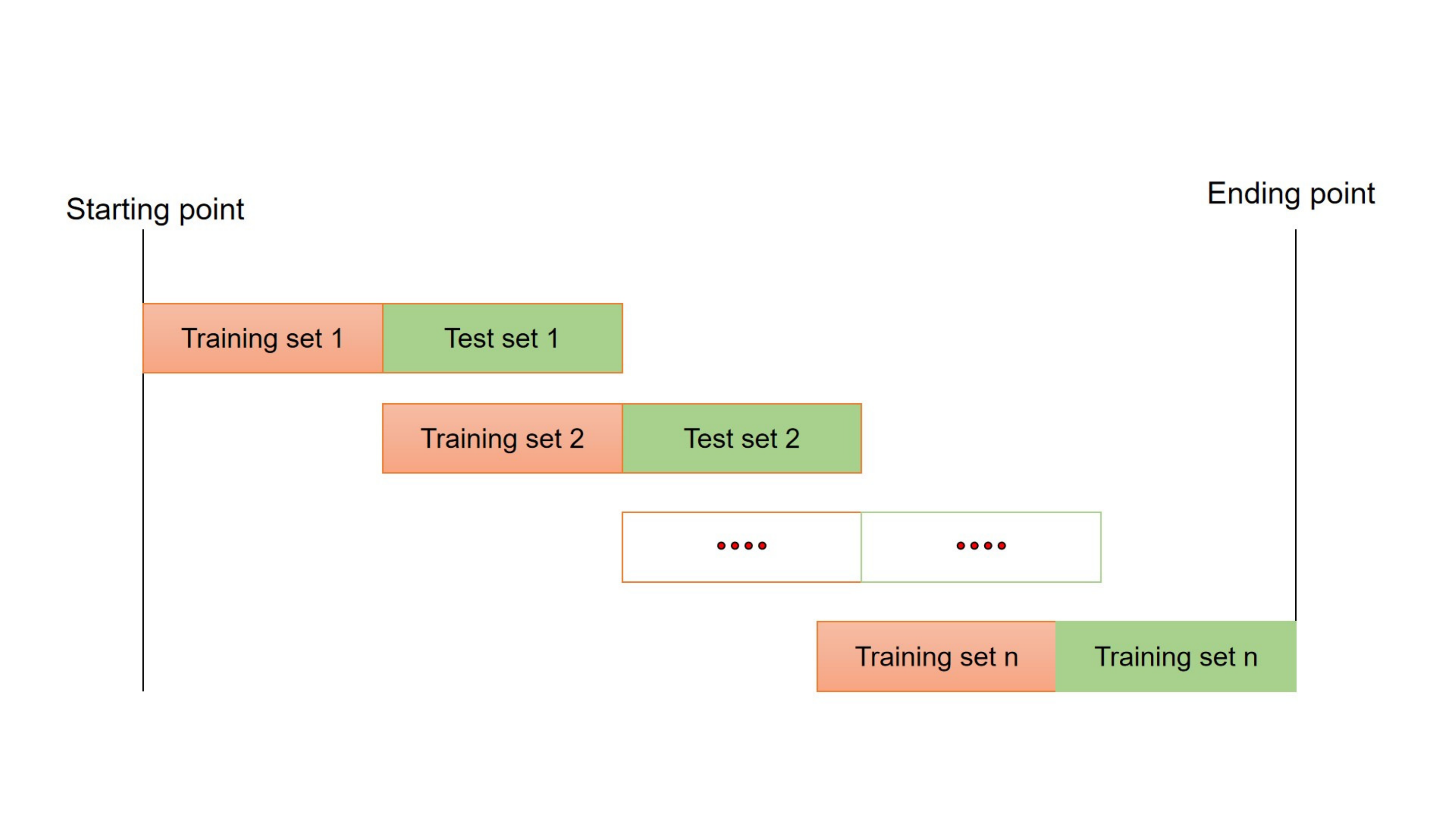}  
  \caption{The rolling training and test processes of the RRL algorithm for portfolio trading.}
  \label{f1}
\end{figure}

\begin{algorithm}
	\caption{Training of the PCA\&DWT RRL.}\label{pc}
	$\textbf{Input}$: $ \bm{X}_t$: Feature matrix; $\bm{r}_t$: Stocks return vector; $T$: Training window size; $\rho$: Learning rate; $\lambda$: Parameter for $\ell_2$ regularization; $\delta$: Transaction cost rate; $N$: The number of epochs; $\kappa$: Random seed for generation of initial portfolio weight; $\epsilon$: Iteration stopping threshold;  \\
	$\textbf{Output}$: $\bm{\Theta}_n^\ast$: Optimized system parameter matrix; \\
	$\textbf{Procedure}$:
	Initialization: $\bm{\Theta}_0 \leftarrow \mathcal{N}(0,1)$, $\bm{F}_0= \bm{0}$, $\frac{\partial\bm{ F}_0}{\partial \bm{\Theta}_0}=\bm{0}$, $W_0 = 1$; \\
	\For{$ n = 0,1,...,N-1 $}{
		\For{$t = 1,..., T-1 $} {
			Receive feature matrix: $\bm{X}_t$, stocks return vector: $\bm{r}_t$;\\
			Calculate $\bm{Y}_t = (\bm{X}_t \otimes \bm{\Theta}_n)\cdot \bm{1}$; \\ 
			Calculate $\bm{f}_t = \tanh(\bm{Y}_t)$; \\
			Calculate $\bm{F}_t =$ softmax($\bm{f}_t$); \\
			Calculate $R_t$, $\frac{\partial R_t}{\partial \bm{F}_t}$, $\frac{\partial R_t}{\partial \bm{F}_{t-1}}$ and $\frac{\partial \bm{F}_t}{\partial\bm{\Theta}_n}$, respectively;
		}
		Calculate Sharpe ratio: $S_T^n$; \\
		\If {$n \geq 2$ \, \text{and} \, $|{S_T^{n}-S_T^{n-1}}| \leq \epsilon$}{
			Stop iteration;
		}		
		Calculate $\frac{dS_T^n}{d\bm{\Theta}_n}$; \\
		Update weight according to:
		$\bm{\Theta}_{n+1} = (1-\rho\cdot\lambda)\cdot\bm{\Theta}_n + \rho\cdot\frac{dS_T^n}{d\bm{\Theta}_n} $;
		
	}
	
\end{algorithm}
Based on what we discussed before, we design Algorithm \ref{pc} to train our model, which aims to obtain the optimized system parameter matrix $\bm{\Theta}_n^\ast$. Then, $\bm{\Theta}_n^\ast$ is directly applied to the same algorithm within one training epoch with the test/trading window size $M$, which gives us the out-of-sample result of the PCA\&DWT RRL method. Afterwards, the training and trading processes are repeated forward until the last batch of trading periods. A graphical representation of the rolling training and test is given in Figure \ref{f1}. When training, the agent is trained for $N$ epochs and the process is early stopped if the objective value is not improved in two consecutive epochs. This is also to avoid overfitting of the RRL algorithm and ensure a better generalization ability of the algorithm. Note the value of $T$ and $M$ can be fine-tuned to fit the market structure underlying different stock price patterns, which should improve the performance theoretically, if there is a similar market pattern in the rolling training and trading windows.  



\section{Experiments}
\subsection{Data Sets}
We test the proposed portfolio trading system on real data sets composed of different number of stocks. As to the number of stocks in a portfolio, we notice that many studies considered a portfolio with less than ten stocks \cite{2021Mean,2020Portfolio,almahdi2017adaptive} and holding too many different stocks is tough to manage for individual investors. \cite{2018Decision} argued that a portfolio with 7 assets is more appropriate than others for portfolio selection under the machine learning context. Therefore, we construct portfolios with cardinality $k=4, 5, 6, 7, 8$ using different stocks as different data sets which are then inputted to the trading algorithm to uncover sequential portfolio weights. Specifically, we randomly choose 8 different stocks with ticker symbols XOM, VZ, NKE, AMAT, MCD, MSFT, AAP and NOV from S\&P500 index which is arguably representative of the general stock market condition in the US, then we form portfolios with different cardinalities using these stocks. Note that due to the RRL trading algorithm where Jacobian matrices are calculated, the order of different stocks in a portfolio matters. However, our empirical results find that the effect of stock orders is not significant. Therefore, we form portfolios with the listed stocks order one by one without loss of generality. For each stock, five data streams are collected from Yahoo Finance \footnote{accessible from \url{https://finance.yahoo.com/}.}, which consist of the daily prices (Open, High, Low, Close) and Volume over the period of 31/12/2009 to 29/12/2017. 
We allow each stock price series to exhibit a unique behavior or pattern to ensure the universality of the data sets. Table \ref{t2} exhibits the summary statistics of the close prices for the 8 stocks.  
\begin{table}
  \caption{Summary statistics for the selected stocks (\$)}
  \label{t2}
  \begin{tabular}{cccccl}
    \toprule
    Symbols & Mean & Std. & Max. & Min. & Range\\
    \midrule
    XOM & 83.73 & 9.66 & 104.38&  56.57 & 47.81 \\
    VZ & 44.02 & 7.39 & 56.53 & 25.26 & 31.27  \\
    NKE & 38.18 & 15.37 & 67.17  & 15.33  & 51.84 \\
    AMAT & 20.26 & 10.69  & 58.80  & 9.85  &  48.95\\
    MCD & 101.25& 23.53 & 174.20 & 61.45 & 112.75 \\
    MSFT & 41.47 & 15.52 & 86.85 & 23.01 & 63.84 \\
    AAP & 106.08 & 42.34  & 200.38 & 39.16 & 161.22 \\
    NOV & 53.19 & 16.28 & 86.43 & 26.34& 60.09 \\
  \bottomrule
\end{tabular}
\end{table}

\subsection{Performance Metrics}
The metrics used to measure the performance of the proposed portfolio trading system in the real financial market are: Net Profit (NP) which is final wealth $W_T$ accumulated by the RRL trading over all the trading periods minus the initial wealth $W_0$; Annualized Percentage Yield (APY), representing the annualized percentage gain; Annualized Sharpe Ratio (ASR), representing the annualized risk adjusted return and we assume the risk-free return is 4\% per annum and there are 252 trading days each year; Maximum Drawdown (MDD) which measures the profit decline percentage from peak value (PV) before largest drop and lowest value (LV) before new high established of an investment during a specific period; Calmar Ratio (CR), indicating the level of risk taken to achieve a return and a higher CR suggests that the system's return is not at the risk of large drawdowns and vice verse. The higher the numerical value of all these metrics are, the better the performance of the trading system is except for MDD for which a lower value is preferred, since most investors are risk-averse. The definitions for these metrics are summarized in Table \ref{t3}.
 \begin{table}[t]        
    \caption{Metrics used to measure the performance of different trading systems}
     \label{t3}
        \begin{tabular}{ccccc}
            \toprule
             NP &  APY & ASR &  MDD & CR  \\  
             \midrule
             $W_T-W_0$ & $(\frac{W_T}{W_0})^{\frac{252}{T}} - 1$  & $\frac{\text{APY} - 0.04}{\text{Std}(R_{1,\ldots,T})\cdot \sqrt{252}}$  & $\frac{\text{PV}-\text{LV}}{\text{PV}}$ & $\frac{\text{APY}}{\text{MDD}}$ \\
            \bottomrule
        \end{tabular}

    \end{table}
\subsection{Benchmark Strategies}
A way to show the efficiency of the proposed portfolio trading system would be to compare the performance of the system with other benchmark methods. In this paper, we compare the proposed trading method with the baseline, Uniform Constant Rebalanced Portfolios (UCRP), which rebalances to a uniform portfolio with equal weight of each stock at the beginning of every period \cite{li2015moving}; OLMAR, which is representative of the notable online portfolio selection techniques in recent years \cite{li2015moving}; the mean-variance portfolio selection model (MV), which utilizes the Monte Carlo method to generate different portfolios, that is, randomly create a set of weights and calculate the mean and variance of each portfolio under the weight, then choose the corresponding weight with the highest Sharpe ratio to allocate wealth for each period over the horizon \cite{2021Mean,2020Portfolio,2018Decision,almahdi2017adaptive}. Note that the mean-variance portfolio choice theory \cite{markowitz1952portfolio} usually aims for single period portfolio selection, researchers sometimes assume that the obtained portfolio weight is optimal for the entire investment horizon in hindsight, which may be not true. We also compare the proposed method with the original RRL trading method (LAG RRL) which simply uses the lagged historical daily return of each stock as features plus a RRL trading module \cite{moody2001learning}. Furthermore, in order to evaluate the effect of using PCA and DWT techniques, we also present the results of the proposed method without PCA and DWT layers (TA RRL). 

\subsection{Hyperparameters}
A common feature of most machine learning models is that their performance highly depends on the setting of hyperparameters which are parameters set before the training process begins. In our case, the performance of the trading system is similarly affected by hyperparameters from each module of the system. Besides the hyparameters described before in the data preprocessing module of the system, hyperparameters of the RRL trading module also matter. 
Furthermore, theoretically speaking, for each feature set, there is an optimal set of hyperparameters associated.
However, due to multiple factors, such as the big number of hyperparameters, the large value space of each hyperparameter and the interdependence amongst different hyperparameters, determination of the optimal set of hyperparameters is almost impossible. Even if one could find the optimal set of hyperparameters for one data set, these hyperparameters is most likely to be sub-optimal for other data sets. Therefore, in this paper, the value of hyperparameters of the proposed portfolio trading system is set empirically to try to ensure an overall good performance on all data sets, unless otherwise stated. Specifically, besides the hyperparameter in the data preprocessing module, we empirically fix $T=100, \rho=0.1, \lambda=0.01, \delta=0.001, N=100, \kappa=42, \epsilon=0$ and the rolling trading window size $M=100$. Note that in terms of the transaction cost, this paper only considers brokerage cost as it is directly controlled by individual investors. Referring to parameter setting of several empirical research \cite{2020Portfolio,almahdi2017adaptive}, we decide to simulate the transaction cost as $\delta=0.001$ or $10$ bps without loss of generality for all listed strategies. Moverover, to ensure a fair comparison, we make the shared hyperparameters same for three RRL based strategies and the LAG RRL is additionally fine-tuned in its lagged length of return series, while OLMAR is also fine-tuned in its hyperparameters, reversion threshold and look-back window and we use Monte Carlo method to simulate 50,000 different sets of portfolio weights for the MV strategy, which would arguably cover most possible portfolios, from statistical point of view \cite{2021Mean,2020Portfolio}.

\subsection{Numerical Results}
\begin{table}
  \caption{Numerical value of the performance of different strategies on different portfolios }
  \label{t4}
  \resizebox{\columnwidth}{!}{
  \begin{tabular}{cccccccl}
    \toprule
    & k  & UCRP & OLMAR & MV  & LAG RRL & TA RRL & PCA\&DWT RRL \\
    \midrule
    NP &&&&&&& \\
    &4 & 1.64 & 0.64&2.32&0.97&1.44& \textbf{3.17} \\
    & 5 & 1.60&0.54&1.87&0.48&1.02& \textbf{3.07} \\
    & 6 & 1.73&0.80&1.80&0.50&1.03&  \textbf{3.02} \\
    & 7 & 1.62&1.19&1.75&0.55&0.86&  \textbf{2.80} \\
    & 8 & 1.40&1.32&1.80&0.35&0.93& \textbf{2.43} \\
    APY & & & & & & & \\
     &4 & 0.15&0.07&0.19&0.10&0.13& \textbf{0.22} \\
    & 5 & 0.15&0.06&0.16&0.06&0.10& \textbf{0.22} \\
    & 6 & 0.15&0.09&0.16&0.06&0.10& \textbf{0.21} \\
    & 7 & 0.15&0.12&0.16&0.06&0.09& \textbf{0.21} \\
     &8 & 0.13&0.13&0.16&0.04&0.10& \textbf{0.19} \\
    ASR &&&&&&& \\
    & 4& 0.71&0.19&0.85&0.38&0.59& \textbf{1.10} \\
    & 5& 0.76&0.14&0.88&0.11&0.44& \textbf{1.21} \\
    & 6 &0.81&0.30&0.87&0.13&0.45& \textbf{1.22} \\
    & 7 & 0.76&0.49&0.86&0.17&0.35& \textbf{1.14} \\
    & 8 & 0.63&0.54&0.85&0.02&0.38& \textbf{0.98} \\
    MDD &&&&&&& \\
    &4 & 0.20&0.25&0.19&0.22&0.16& \textbf{0.13} \\
    &5 & 0.17& 0.26& \textbf{0.14} &0.20&0.17& 0.15 \\
    &6 & 0.17&0.24&0.13&0.18&0.19& \textbf{0.12} \\
    &7 & 0.14&0.24&0.13&0.16&0.16& \textbf{0.11} \\
    &8 & 0.17&0.25&0.13&0.23&0.18& \textbf{0.12} \\
    CR &&&&&&& \\
    &4 & 0.76&0.29&1.00&0.45&0.82& \textbf{1.66} \\
    &5 & 0.85&0.24&1.20&0.28&0.61& \textbf{1.49} \\
    &6 & 0.93&0.36&1.21&0.32&0.55& \textbf{1.85} \\
    &7 & 1.06&0.50&1.18&0.40&0.57& \textbf{1.94} \\
    &8 & 0.77&0.52&1.24&0.19&0.53& \textbf{1.61} \\
     \bottomrule
  \end{tabular}
  }
\end{table}
\begin{figure}[h]
  \centering
  \includegraphics[height=12cm, width=\columnwidth]{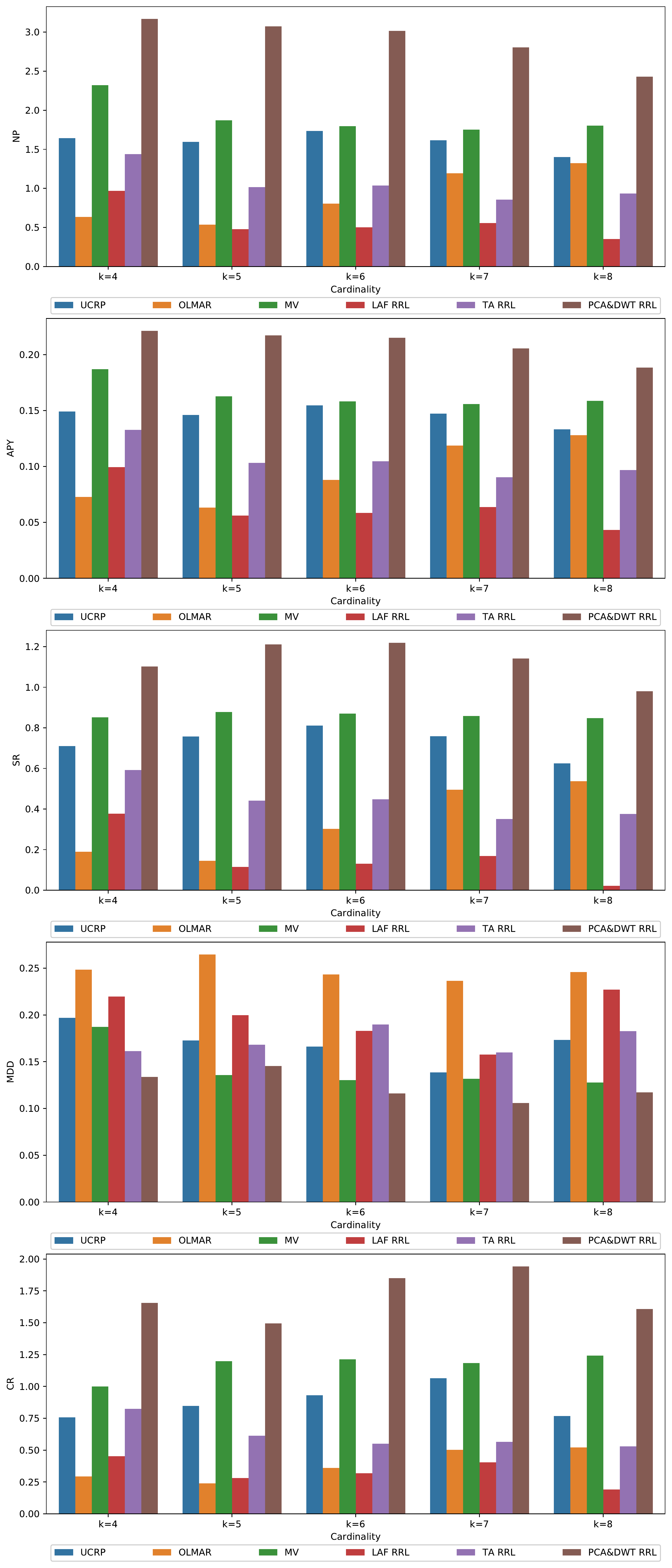} 
  \caption{Boxplots of the performance of different strategies on various portfolios.}
  \label{f2}
  \Description{Each panel shows boxplots of numerical results of different strategies on portfolios with various cardinalities.}
\end{figure}

\begin{figure}
  \centering
  \includegraphics[width=\columnwidth]{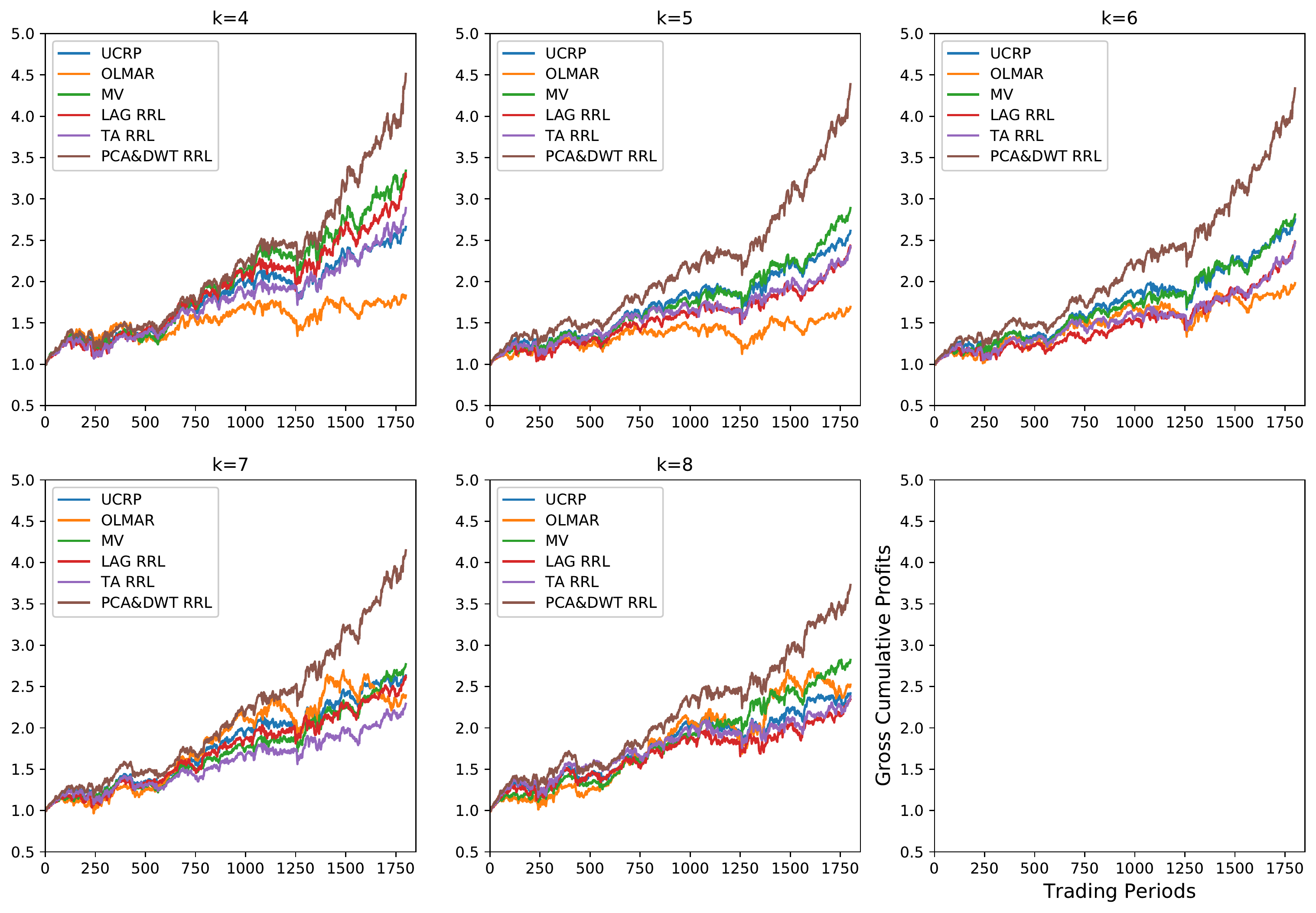}  
  \caption{Gross cumulative profits of different strategies on various portfolios with $\delta=0$. Repetitive labels are placed in the last empty panel.}
  \label{f3}
  \Description{Each subplot represents the cumulative gross profit over the horizon obtained by different strategies on portfolios with various cardinalities under the transaction cost 0.}
\end{figure}
\begin{figure}
  \centering
  \includegraphics[width=\columnwidth]{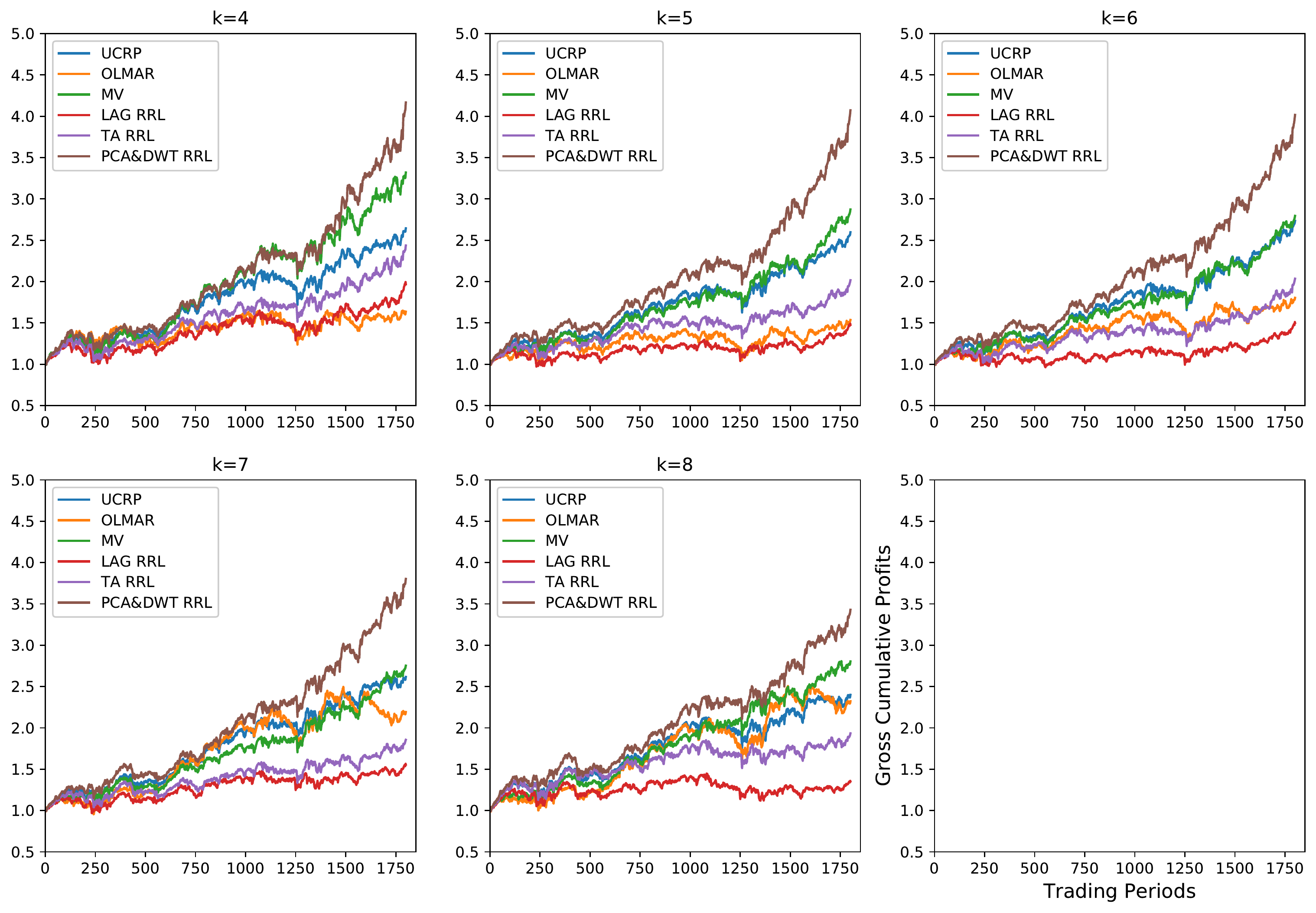}  
  \caption{Gross cumulative profits of different strategies on various portfolios with $\delta=10$ bps. Repetitive labels are placed in the last empty panel.}
  \label{f4}
  \Description{Each subplot represents the cumulative gross profit over the horizon obtained by different strategies on portfolios with various cardinalities under the transaction cost 10 bps per share.}
\end{figure}
\begin{figure}
  \centering
  \includegraphics[width=\columnwidth]{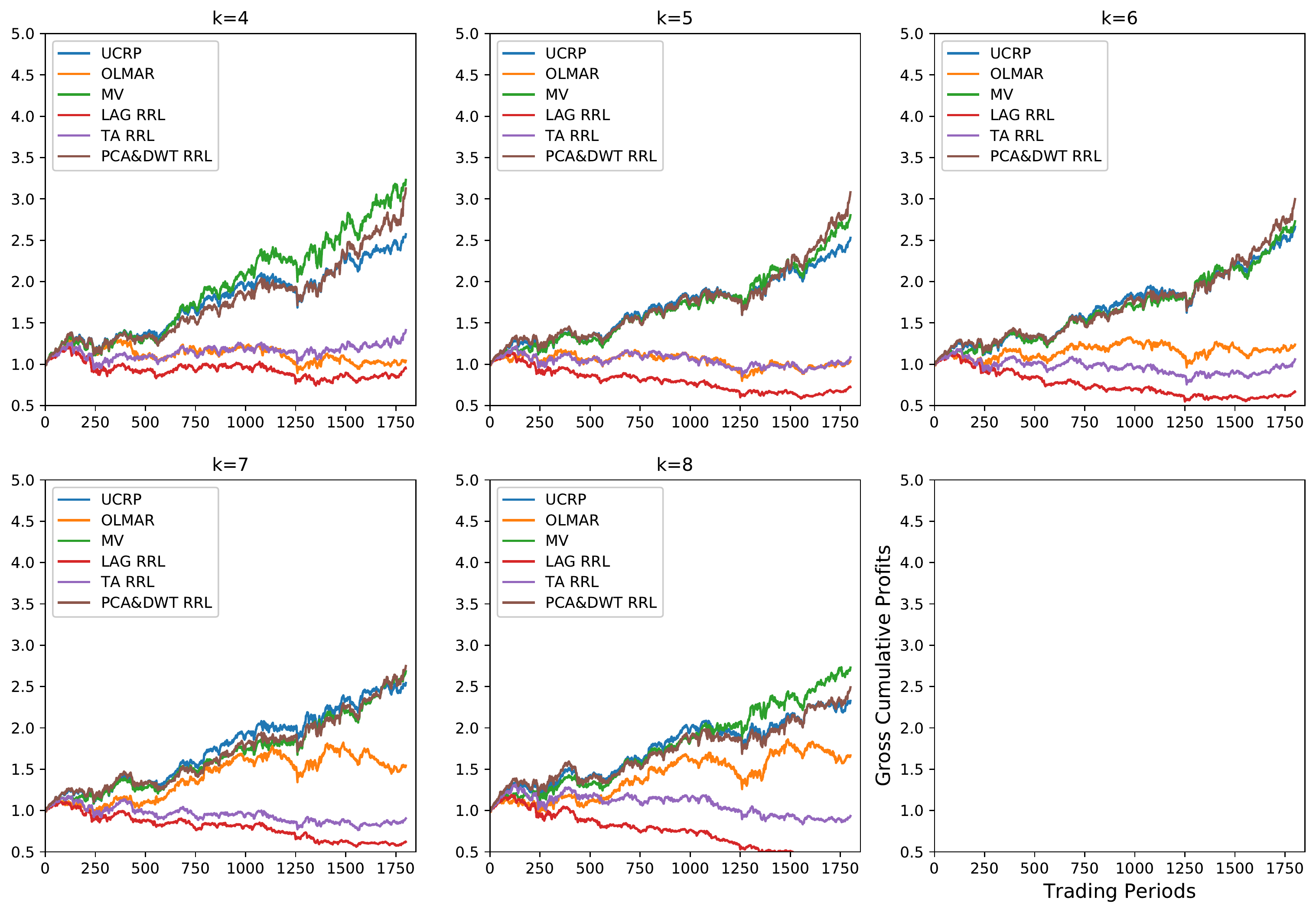} 
  \caption{Gross cumulative profits of different strategies on various portfolios with $\delta=50$ bps. Repetitive labels are placed in the last empty panel.}
  \label{f5}
  \Description{Each subplot represents the cumulative gross profit over the horizon obtained by different strategies on portfolios with various cardinalities under the transaction cost 50 bps per share.}
\end{figure}

While UCRP and MV are passive investment strategies whose profits are mostly determined by the movement of the underlying assets' price, the other methods are active in taking actions in markets.   
The numerical results of all strategies' performance are presented in Table \ref{t4} with the best results obtained by all strategies on each portfolio stressed in bold face. Additionally, Figure \ref{f2} presents these numerical results with boxplots. 

By examining the obtained results, it is apparent that the proposed trading system clearly outperforms all other strategies in term of almost all metrics on portfolios with different cardinalities. Specifically, the PCA\&DWT RRL earns the highest NP, APY, ASR, CR and lowest MDD, except for the MDD of the portfolio containing 5 stocks, than others on various portfolios, which shows that this portfolio trading system is not only effective but superior to the benchmark in most cases. Secondly, PCA\&DWT RRL substantially outperforms other two RRL based strategies, meaning that combining PCA and DWT techniques implemented on technical indicators plays an important role in improving the performance of the trading system. Especially, feature preprocessing is vital for RRL trading algorithm. Thirdly, it is found that UCRP and MV are relatively good strategies except PCA\&DWT RRL, revealing following the market is always a viable strategy, while OLMAR is less competitive on these data sets, which seems to verify that OLMAR is more adapted to large portfolios management.

Finally, we test the performance of all strategies w.r.t. transaction costs. Given many online brokers provide free stock trading, we set $\delta$ as 0 bps, 10 bps and 50 bps, respectively, per share for more general settings. Figure \ref{f3} presents the trend of gross cumulative profits $W_T$ of different strategies w.r.t. various portfolios with $\delta=0$, while Figure \ref{f4} and \ref{f5} reveal the performance with $\delta=$ 30 bps and 50 bps, respectively. It is clear that the proposed system is negatively affected by transaction costs, since on each particular portfolio, the higher the transaction cost rate is, the lower the final cumulative profit obtained by PCA\&DWT RRL strategy. On the contrary, passive investment strategies, UCRP and MV, are less affected by transaction costs, since they just need to rebalance the portfolio affected by the underlying stocks price movement which is normally tiny. However, on most portfolios, especially when the transaction cost rates are low, PCA\&DWT RRL system significantly outperforms all other benchmark strategies in gross profits, which reveals that the proposed trading system is robust, consistent and competitive in making profits compared to other strategies. Nevertheless, the proposed strategy seems to be unsuitable for large portfolios trading since all figures show that the more stocks contained in a portfolio, the less the profit gained by our strategy with other hyperparameters fixed.

\section{Conclusion}
This paper proposes a novel portfolio trading system, PCA\&DWT RRL, which not only embeds the portfolio rebalance function into the algorithm to trade portfolios at each period directly, but also combines PCA and DWT to preprocess the technical indicators extracted from the original stock price and volume data. The experimental results demonstrate that the proposed system consistently outperforms other portfolio selection strategies from previous literature. Moreover, we find that feature preprocessing is vital for the RRL trading algorithm in this setting. Future research could exploit more about the portfolio rebalance function to further improve the performance and make the system more adapted to large-scale portfolio selections for financial institutions.






\bibliographystyle{ACM-Reference-Format}
\bibliography{sample-base}






\end{document}